\documentclass[jkps,preprint,showpacs,showkeys]{revtex4}
\usepackage{graphicx}
\usepackage{amssymb}
\usepackage{bm}

\begin{document}
\setcounter{page}{1}
\title[]{Newton-Wigner position operator and its corresponding spin operator in relativistic quantum mechanics}
\author{Taeseung \surname{Choi}}
\email{tschoi@swu.ac.kr}
\thanks{Fax: +82-2-970-5907}
\affiliation{Division of General Education, Seoul Women's University, Seoul 139-774}
\affiliation{School of Computational Sciences, Korea Institute for Advanced Study, Seoul 130-012, Korea }
\date[]{}

\renewcommand{\abstractname}{}
\begin{abstract}
A relativistic spin operator is to be the difference between the total and orbital angular momentum. 
As the unique position operator for a localized state, 
the remarkable Newton-Wigner position operator, which has all 
desirable commutation relations as a position operator, can give a proper spin operator. 
Historically important three spin operators respectively proposed by Bogolubov et al., Pryce, 
and Foldy-Woutheysen are investigated to manifest a corresponding spin operator to the 
Newton-Wigner position operator. 
We clarify a unique spin operator in relativistic quantum mechanics described by the Dirac Hamiltonian. 

\end{abstract}

%Use showkeys class option if keyword
\maketitle
%\maketitle

\section{Introduction}

%However, the physically intuitive understanding of the spin operator can be given by considering the 
%historical approach in relativistic quantum mechanics. 
%Moreover, the relations of previous major works are not fully appreciated in most recent work \cite{Bauke}. 
According to very recent works that have introduced many spin operators suggested in research papers 
\cite{Bauke14, Caban13}, there is still lack of consensus on a 
	proper relativistic spin operator for a massive spin $1/2$ particle. In non-relativistic quantum 
	mechanics spin angular momentum can be defined by the difference between total angular momentum 
	and orbital angular momentum. 
		This definition of the spin operator is also valid for relativistic quantum mechanics. 
		In this approach, by definition, there is a key prior step to find a proper relativistic spin operator, 
		i.e., finding a proper position operator for relativistic systems \cite{Pryce48, NW49, FW50, Jordan63, Fleming64}. 
		In non-relativistic quantum mechanics, the Pauli spin operator $\boldsymbol{\sigma}/2$ 
		is related with the usual position operator ${\bf x}$ by the formula 
$\boldsymbol{\sigma}/2= {\bf J}- {\bf x}\times {\bf P}$, where ${\bf J}$ is a total angular momentum 
and ${\bf P}$ is a momentum. However, a relativistic spin would not be a Pauli spin operator 
$\boldsymbol{\sigma}/2$ for a massive spin $1/2$ particle, 
where $\boldsymbol{\sigma}$ is a spatial three-vector with the usual Pauli matrices as components, 
because the usual position operator ${\bf x}$ has 
some serious drawbacks to become a physical position operator of a massive spin $1/2$ particle, whose 
dynamics is governed by the Dirac Hamiltonian. 
For instance, the velocity operator of such a particle, can be called Dirac particle, 
is given by the commutation relation of the usual position operator of the free massive particle 
and the Dirac Hamiltonian, and its eigenvalues are $+ 1$ or $-1$ in the unit $c=1$, which 
implies that the speed of a massive Dirac particle is allowed to be only the speed of light. 
This fact seems to be very weird because the Dirac equation describes all the particles with arbitrary momentum 
from the rest particle to fast relativistic particles. Furthermore, the time derivative of the velocity operator 
is known to show a very rapid oscillating motion that is called Zitterbewegung, 
which means that the eigenvalue of the usual position operator cannot give a good result 
for a position measurement of a massive Dirac particle like an electron. 

Such ambiguities originating from the usual position operator have motivated to divert 
to an average position operator, i.e., a center of distributed mass according to the Zitterbewegung. 
On the contrary to Newtonian mechanics, however, defining a center of mass is highly nontrivial in relativity. 
Thus, Pryce studied possible three mass-center position operators in the names of 
the definitions (c), (d), and (e), and their related spin operators in the Ref. \cite{Pryce48}. 
Later, Newton and Wigner (NW) \cite{NW49} obtained a unique position operator by defining a unique localized states 
formulated on the requirement of natural invariance for elementary systems with positive energy, 
and mentioned that the unique position operator is the same as the Pryce position operator (e) 
for a massive spin-1/2 particle. 
Furthermore, more general considerations \cite{Fleming64, Lorente74} have verified the validness of 
the NW position operator. Then, the Pryce spin operator (e) corresponding to Pryce position operator (e) 
can be a proper spin operator for a massive Dirac particle. However, the Pryce spin operator (e) has not 
been fully appreciated and investigated because another way to explore a proper relativistic spin operator 
has been suggested by several authors in literatures \cite{Bogolubov78, FW50}. 

Other important approaches to obtain a proper relativistic spin operator can be a group theoretical approach 
\cite{Bogolubov78} or a unitarily transformed representation approach \cite{FW50}. 
In a group theoretical approach, Bogolubov et al. started to study a spin operator from a linear combination 
of the Pauli-Lubanski (PL) vector because the inhomogeneous Lorentz (Poincar´e) group has two Casimir invariants and 
one of them is the square of the PL vector responsible for spin. 
In order to determine a spin operator, they used the three physical requirements on the linear combination of 
PL vector and obtained an axial spin three-vector \cite{Bogolubov78}. 
On the other hand, in unitary representation approach, Foldy and Woutheysen introduced a new representation 
(so-called Foldy-Woutheysen (FW) representation) that is equivalent to the standard Dirac representation. 
In this representation, they found that the Hamiltonian becomes diagonal so that the usual forms of the position 
and spin operators satisfy the desirable properties to being the position and spin operators. 
Their usual position and spin operators in the FW representation are shown to be the mean position and 
mean spin operators in the standard Dirac representation.

Although the remarkable NW position operator is believed to be a proper position operator in relativistic quantum mechanics, 
however, its corresponding spin operator has not been considered explicitly to find a proper spin operator 
for a massive spin-1/2 Dirac particle in literatures. 
In this paper, we will explicitly study the relations between the NW position operator and the suggested spin operators in literatures. 
Especially, we will focus on the three historically important spin operators proposed by Bogolubov et al., Pryce and Foldy-Woutheysen. 
The definitions of the proposed relativistic spin operators will be revisited to clarify a unique spin operator in relativistic 
quantum mechanics described by the Dirac Hamiltonian. 
In Sec. II, we will briefly review and discuss the spin operator derived by Bogolubov et al. 
In Sec. III, we will study the position and spin operators proposed by Pryce. 
In Sec. IV, we will discuss the Foldy-Woutheysen spin and position operators with relate to Pryce and 
Newton-Wigner position operators. Our results will be summarized in Sec. V.

%In this paper we will show explicitly the equality of the Pryce spin (e) and the Foldy-Wouthesen (FW) spin operators 
%\cite{Price, FW}. Bauke et al. \cite{Bauke} has dealt with different Pryce spin, however, which one of Pryce spins 
%are used is not clear because they used nonconventional notation as they have noted. 
%The difference from Pryce (e) is definite because Pryce (e) will be shown to be equal to their FW spin, 
%which can be seen in TABLE 1 and FIG 1 of their paper. 
%We would like to note that Chakrabarti spin and FW spin operators are equivalent in the sense that give the same physical 
%results other than FIG 1. As we have noted in JKPS, the different scalar product must be used 
%for Chakrabati spin and FW spin operators, however, to obtain the results of spin expectation values in FIG 1. Bauke et al. 
%has used the smae wave function and scalar product. This misuse of scalar product for Chakrabarti spin makes the 
%results in FIG 1. different from the result of FW.  -> Just simply note!!

%We expect that the eigenvalue of the velocity operator for a positive energy particle becomes ${\bf p}/p^0$, 
%where $p^\mu=(p^0, {\bf p})$ is 4-momentum.  

%Even though a proper relativistic spin has been derived from the general linear combination of 
%Pauli-Lubanski vector, physical interpretation

%In semiclassical understanding, the relativistic position operator becomes nonlocal.

\section{Bogolubov's Spin Operator}
\label{sec:Bog}

Bogolubov et al. have derived the following spin operator 
\begin{eqnarray}
{\bf S}_{\scriptsize{BG}}= \frac{1}{m} \left( {\bf W} - \frac{W^0 {\bf P}}{m+P^0}\right),
\label{eq:SPINBg}
\end{eqnarray} 
starting from a combination of the 
Pauli-Lubanski (PL) vector 
$W^\mu=\frac{1}{2} \epsilon^{\mu\nu \rho\sigma}J_{\nu\rho} P_\sigma$ \cite{Bogolubov78}. 
Here boldface letters are used for denoting contra-variant vectors, and $P^\mu$ and 
$J^{\mu\nu}$ are momentum and Lorentz operators that are the translation and Lorentz transformation 
generators in the Poincar\'e group, respectively. Einstein summation convention is used for 
the Greek letters $\mu=\{0,1,2,3\}$ and will be used for the Latin letters $k=\{1,2,3\}$. 
$\epsilon^{\mu\nu \rho\sigma}$ is a Levi-Civita 
symbol in Minkowski space with $\epsilon^{1230}=1$ and the metric $g^{\mu\nu}=\mbox{diag}(+,-,-,-)$ is 
used. $m$ is the invariant mass of a relativistic particle.

To derive the spin operator ${\bf S}_{\scriptsize{BG}}$ in Eq. (\ref{eq:SPINBg}), Bogolubov et al. used the three 
requirements: 1) $[S_{\scriptsize{BG}}^i, S_{\scriptsize{BG}}^j]=i\epsilon_{ijk}S_{\scriptsize{BG}}^k $ 
(${\bf S}_{\scriptsize{BG}}$ must satisfy the $su(2)$ 
algebra) 2) $[J^i, S_{\scriptsize{BG}}^j]=i\epsilon_{ijk}S_{\scriptsize{BG}}^k $ 
(${\bf S}_{\scriptsize{BG}}$ must be a spatial three-vector) 
3) ${\bf S}_{\scriptsize{BG}}$ must be an axial vector. 
Here $\epsilon_{ijk}$ is a three-dimensional Levi-Civita symbol 
with $\epsilon_{123}=1$. They started with an axial vector 
\begin{eqnarray}
{\bf S}=a \left( {\bf W} - b W^0 {\bf P}\right),
\end{eqnarray}
where $a$ and $b$ are functions of ${\bf P}\cdot {\bf P}=\sqrt{(P^0)^2 +m^2}$.
 The fact that the vector ${\bf S}$ is an axial vector can be easily checked by using the transformations 
$(W^0, {\bf W}) \rightarrow (-W^0, {\bf W})$ and $(P^0, {\bf P})\rightarrow (P^0, -{\bf P})$ under the parity 
(spatial inversion).  

It seems that the spin operator ${\bf S}_{\scriptsize{BG}}$ in Eq. (\ref{eq:SPINBg}) 
could be a spin operator in a moving frame because the spin 
operator ${\bf S}_{\scriptsize{BG}}$ is represented by the 4-vector operators $W^\mu$ and $P^\mu$. 
In fact many authors in literatures have considered the spin operator in Eq. (\ref{eq:SPINBg}) as a relativistic spin 
in a moving frame \cite{Alsing12}. To investigate whether the spin operator ${\bf S}_{\scriptsize{BG}}$ 
could be a spin operator in an arbitrary frame 
specifically, let us consider the Lorentz boost, 
\begin{eqnarray}
\label{eq:SLB}
L({\bf p})^0_{\ 0} = \frac{p^0}{m},~~~ L({\bf p})^0_{\ i} = \frac{p^i}{m}, 
~~~L({\bf p})^i_{\ j} &=& \delta_{ij} + \frac{p^i p^j}{ m(p^0+m)},
\end{eqnarray}
 with the Kronecker-delta $\delta_{ij}$, which transforms the momentum $\tilde{q}^\mu$ of the particle in 
one moving frame $\widetilde{\mathcal{O}}$ to the momentum ${ q}^\mu$ in another moving frame ${\mathcal{O}}$ 
as $q^\mu=L({\bf p})^\mu_{\phantom{\mu}\nu} {\tilde{q}}^\nu$. 
Here we use lower case ${\bf p}$ because the momentum in a specific frame is not an operator. 
However, we use upper case $W$ for PL vector because the context will make the use of the operators or their values clear.  
Then, the PL vector $\widetilde{W}^\mu$ in the frame $\widetilde{\mathcal{O}}$ can be written by using the 
PL vector $W^\mu$ in the frame $\mathcal{O}$, by using 
the inverse Lorentz transformation as
\begin{eqnarray}
\widetilde{W}^i&=& L(-{\bf p})^i_{\phantom{i}0}W^0 + L(-{\bf p})^i_{\phantom{i}{j}}W^j \\ \nonumber
&=& W^i -\frac{W^0}{m+p^0}p^i,
\label{eq:BPLLT}
\end{eqnarray}
because $L({\bf p})^{-1}=L(- {\bf p})$. 
This seems to imply that the spin operator ${\bf S}_{\scriptsize{BG}}$ in Eq. (\ref{eq:SPINBg}) is nothing but
\begin{eqnarray}
{\bf S}_{\scriptsize{BG}}= \frac{\widetilde{\bf W}}{m}.
\label{eq:BSPLT}
\end{eqnarray}
The equality of two Eqs. (\ref{eq:SPINBg}) and (\ref{eq:BSPLT}), however, holds only when the PL vector 
$\widetilde{\bf W}$ in Eq. (\ref{eq:BSPLT}) represents the PL vector in the particle rest frame. 
This is because the PL vector operator ${\bf W}$ and the momentum operator ${\bf P}$ 
in Eq. (\ref{eq:SPINBg}) must have their values, respectively, in the same reference frame. 
This means that the particle has the momentum ${\bf p}$ in the reference frame $\mathcal{O}$ and 
so the reference frame $\widetilde{\mathcal{O}}$ should be the particle rest frame. 
Bogolubov et al. themselves also remarked that the $\widetilde{W}^i$ in Eq. (\ref{eq:BPLLT}) is nothing but 
the spatial component of the 4-vector transformed to the particle rest frame.    

In the standard Dirac representation, $\widetilde{W}^\mu=(0, m {\boldsymbol{\Sigma}}/2)$, where 
$$
\boldsymbol{\Sigma}= \left( \begin{array}{cc} \boldsymbol{\sigma} & 0 \\ 0 & \boldsymbol{\sigma} \end{array} \right)
$$
and $\boldsymbol{\sigma}$ has the usual Pauli matrices as its components. 
Hence, the unique axial spin operator obtained from the Bogolubov et al. is 
the 4-dimesnional Pauli spin operator in the particle rest frame, 
which cannot be a spin operator in a moving frame. 
The corresponding position operator is meaningless in the particle rest frame because of the uncertainty 
relation. 
 
Note that the operators in all equivalent representations can be obtained by a similarity transformation 
of the operators in one representation. Hence it is enough to consider the standard Dirac representation 
in defining the position and spin operators.

\section{Pryce Spin Operators}
\label{sec:Pryce}

Pryce has obtained the three position and spin operators explicitly by considering the relativistic generalization 
of the mass-centers defined in Newtonian mechanics \cite{Pryce48}. Recently, in Ref. \cite{Alsing12}, the authors 
have used the generalized definitions for the position and the spin operators of Pryce (c) and (e), 
which use the covariant generators of the Poincar\'e group including the PL vector. 
The generalization allows the definitions of the position and the spin operators being naturally used 
in other representations. 

As is well-known, the spin operator should be given through a second Casimir invariant, 
being the square of the PL vector, of the Poincare group \cite{Ryder1}. 
Hence a general form of spin operators and its corresponding position operators 
seem to be obtained by using the generators of the Poincar\'e group including the PL vector. 
The definitions of the position and the spin operators used in the recent works \cite{Alsing12} are 
\begin{eqnarray}
{\bf R}_{\scriptsize{CM}} &=& -\frac{1}{2} \left( \frac{1}{P^0} {\bf K} + {\bf K} \frac{1}{P^0}\right), ~~
{\bf S}_{\scriptsize{CM}} = \frac{\bf W}{P^0} 
\label{eq:CM} \\
{\bf R}_{\scriptsize{NW}} &=&  {\bf R}_{\scriptsize{CM}}  - \frac{{\bf P}\times {\bf W}}{mP^0(m+P^0)}, ~~
{\bf S}_{\scriptsize{NW}} = \frac{1}{m} \left( {\bf W}- \frac{W^0 {\bf P}}{m+P^0}\right),
\label{eq:NW} 
\end{eqnarray}
where $K^i=J^{0i}$ is the Lorentz boost operator. 
${\bf R}_{\scriptsize{CM}} $ and ${\bf R}_{\scriptsize{NW}}$ are used as generalizations of the mass-center operator 
of Pryce (c) and the NW position operator, respectively. 
Newton and Wigner remarked that the NW position operator is the same as the position operator of Pryce (e) for a 
spin $1/2$ massive particle \cite{NW49}.

%The definitions of Pryce (c) and (e) are 
%\begin{eqnarray}
%{\bf q}_{\scriptsize{(c)}} &=& -\frac{1}{2} \left(\frac{1}{\mathcal{H}_{\scriptsize{D}}} {\bf K} + 
%{\bf K} \frac{1}{\mathcal{H}_{\scriptsize{D}}}\right), ~~
%{\bf S}_{\scriptsize{(c)}} = {\bf J}- {\bf q}_{\scriptsize{(c)}} \times {\bf P} \\ \nonumber
%{\bf q}_{\scriptsize{(e)}} &=& {\bf q}_{\scriptsize{(c)}} +\frac{m(m+E)} 
%{\bf S}_{\scriptsize{(c)}}\times {\bf P},
%\end{eqnarray}
%where 
We will study whether the CM and NW operators in Eqs. (\ref{eq:CM}) and (\ref{eq:NW}) give 
the same results as the Pryce (c) and (e) operators in Ref. \cite{Pryce48}, respectively, 
in the standard Dirac representation because Pryce has obtained the position and the spin operators 
by using the standrad Dirac Hamiltonian.  
To compare the two kinds of operators, we find the explicit expressions for the position and the spin operators 
of Pryce (c) and (e) by using the Poincar\'e generators. 
The position and the spin operators of Pryce (c) and (e) are respectively given as 
\begin{eqnarray}
\label{eq:PRYC}
{\bf q}_{\scriptsize{(c)}} &=& {\bf x} +\frac{1}{(P^0)^2} \left( \frac{1}{m}{\bf P}\times {\bf W}
 + i\gamma^0 \gamma^5 \left( {\bf W}- \frac{W^0 {\bf P}}{m+P^0}\right) \right) \\ \nonumber
{\bf S}_{\scriptsize{(c)}} &=& \frac{1}{(P^0)^2} \left( m {\bf W} + \frac{P^0 W^0 {\bf P}}{m+P^0} 
- {i \gamma^0 \gamma^5 } {\bf W}\times {\bf P}\right), 
 \\ \label{eq:PRYE}
%{\bf S}_{\mbox{(d)}} &=& \frac{1}{m}\left( {\bf W}- \frac{W^0 {\bf P}}{m+P^0}\right) 
%-\frac{i \gamma^0 \gamma^5}{m^2}{\bf W}\times {\bf P} \\
{\bf q}_{\scriptsize{(e)}} &=& {\bf x} +\frac{1}{m P^0} \left(  \frac{1}{m+P^0} {\bf P}\times {\bf W}
+ i \gamma^0 \gamma^5 \left( {\bf W} - \frac{W^0 {\bf P}}{P^0} \right) \right) \\ \nonumber
{\bf S}_{\scriptsize{(e)}} &=& \frac{\bf W}{P^0}- \frac{i\gamma^5 \gamma^0 }{m P^0} {\bf W}\times {\bf P},
\end{eqnarray} 
where ${\bf x}$ is the usual position operator in the standard Dirac representation and 
$$
\gamma^0=\left( \begin{array}{cc} I_2 & 0 \\ 0 & -I_2 \end{array} \right), ~~~
\gamma^5=\left( \begin{array}{cc} 0 & I_2 \\ I_2 & 0 \end{array} \right).
$$
$I_2$ is a two-dimensional identity matrix. 
To express the position and the spin operators with PL vectors, 
we used the following relations in the standard Dirac representation, 
\begin{eqnarray}
\frac{\boldsymbol{\Sigma}}{2} = \frac{\bf W}{m} -\frac{W^0 {\bf P}}{m(m+P^0)} \mbox{ and }
\frac{ \boldsymbol{\Sigma} \cdot {\bf P}}{2}= W^0.
\end{eqnarray} 
One can easily check that the position and the spin operators in Eqs. (\ref{eq:CM}) 
and (\ref{eq:NW}) differ from the position and the spin operators in 
Eqs. (\ref{eq:PRYC}) and (\ref{eq:PRYE}). 
%We will show that the position and spin operators in Eqs. (\ref{eq:CM}) and (\ref{eq:NW}) 
%are not valid when the Dirac Hamiltonian is used for a relativistic particle.  
The basic reason for this discrepancy is why the Dirac Hamiltonian 
$\mathcal{H}_{\scriptsize{D}}$ is not the quantum operator 
corresponding to the $0$th-component of $4$-momentum, $P^0$. 
We will show the details in the below. 
 
Let us first consider the case of Pryce (c) in which 
the position operator ${\bf q}_{\scriptsize{(c)}}$ is defined as
\begin{eqnarray}
{\bf q}_{\scriptsize{(c)}} =-\frac{1}{2}\left( \frac{1}{\mathcal{H}_{\scriptsize{D}}} {\bf K}
+{\bf K} \frac{1}{\mathcal{H}_{\scriptsize{D}}}\right).
\end{eqnarray}
Note that the only difference of the definition of the position operators ${\bf q}_{\scriptsize{(c)}}$ from 
the definition of the position operator ${\bf R}_{\scriptsize{CM}}$ is the replacement of $P^0$ with 
$\mathcal{H}_{\scriptsize{D}}$. 
The Dirac Hamiltonian $\mathcal{H}_{\scriptsize{D}}$ 
is not covariant in the sense that the Dirac Hamiltonian in the moving frame, 
$\boldsymbol{\alpha} \cdot {\bf p}+\beta m$ 
cannot be transformed from the Dirac Hamiltonian in the particle rest frame, $k^\mu=(\beta m, {\bf 0})$, through 
the Lorentz transformation, i.e.,
\begin{eqnarray}
\boldsymbol{\alpha} \cdot {\bf p}+\beta m \neq L({\bf p})^0_{\ \mu}k^\mu,
\end{eqnarray}
where $L({\bf p})$ is the Lorentz boost in Eq. (\ref{eq:SLB}).
Here the Dirac matrices are $\boldsymbol{\alpha}=\gamma^5 \boldsymbol{\Sigma}$ 
and $\beta=\gamma^0$. 
On the other hand, the operator $P^0$ is a covariant operator. 
This implies that the two operators $\mathcal{H}_{\scriptsize{D}}$ and $P^0$ are different each other. 
One can check the difference explicitly in the following commutation relations
\begin{eqnarray}
[{\bf x}, \mathcal{H}_{\scriptsize{D}}] = i \boldsymbol{\alpha}, ~~~
[{\bf x}, P^0]=i\frac{\bf P}{P^0}.
\end{eqnarray}
Note that $\boldsymbol{\alpha} \neq {\bf P}/P^0$. 
The difference of the two commutators ensures the difference of two operators 
$\mathcal{H}_{\scriptsize{D}}$ and $P^0$.

The position operator of Pryce (c) is defined in a particular reference frame 
as the mean of the coordinates weighted with the mass density. To obtain the mass density in a particular 
reference frame the explicit form of the Hamiltonian of the system is needed. 
Hence to obtain the position operator corresponding to the mass-center, 
the definition of center of mass in Eq. (\ref{eq:CM}) should not be used for the definition of 
Pryce (c) in Eq. (\ref{eq:PRYC}). 
This implies that the spin operator in Eq. (\ref{eq:CM}) is also not valid for a massive 
Dirac particle.  

Now, let us consider the case of Pryce (e).
The original definition of the position operator of Pryce (e) is given as
\begin{eqnarray}
{\bf q}_{\scriptsize{(e)}} ={\bf q}_{\scriptsize{(c)}} +\frac{1}{m(m+E)}
{\bf S}_{\scriptsize{(c)}}\times {\bf P} ,
\label{eq:PryceE}
\end{eqnarray} 
where $E=\sqrt{{\bf p}\cdot {\bf p}+m^2}$ is the energy of the particle. 
By substituting the position operators 
${\bf R}_{\scriptsize{CM}}$ for ${\bf q}_{\scriptsize{(c)}}$ in Eq. (\ref{eq:PryceE}), 
the definition of ${\bf q}_{\scriptsize{(e)}}$ becomes 
\begin{eqnarray}
{\bf q}_{\scriptsize{(e)}}= {\bf R}_{\scriptsize{CM}} +\frac{1}{m(m+E)}
{\bf S}_{\scriptsize{(c)}}\times {\bf P}.
\end{eqnarray}
This changes to the definition of 
the NW position operator ${\bf R}_{\scriptsize{NM}}$ in Eq. (\ref{eq:NW}) if 
the relation ${\bf S}_{\scriptsize{CM}} = {\bf W}/{P^0} $ is used. 
The relation ${\bf S}= {\bf W}/P^0$, however, is not invalid for a 
Dirac particle because $P^0$ is used instead of $\mathcal{H}_{\scriptsize{D}}$. 
This means that the NW position operator ${\bf R}_{\scriptsize{NW}}$ and 
the spin operator ${\bf S}_{\scriptsize{NW}}$ are also not valid, similarly to 
the position operator ${\bf R}_{\scriptsize{CM}}$ and the spin operator ${\bf S}_{\scriptsize{CM}}$ 
for a Dirac particle.

The mass-center operator ${\bf q}_{\scriptsize{(c)}}$ cannot be a local position operator because 
its components do not commute each other unlike 
the components of the usual position operator ${\bf x}$. 
The mass-center operator ${\bf q}_{\scriptsize{(e)}}$ is the 
modified one from  ${\bf q}_{\scriptsize{(c)}}$ in order to satisfy the following commutation 
relations  
\begin{eqnarray}
[q^i_{\scriptsize{(e)}}, q^j_{\scriptsize{(e)}}]=0.
\end{eqnarray}
Then the spin operator ${\bf S}_{\scriptsize{(e)}} $ automatically satisfies $su(2)$ algebra. 
Note that $[S^i_{\scriptsize{(c)}}, S^j_{\scriptsize{(c)}}]= m^2 S^k_{\scriptsize{(c)}}/(P^0)^2$.  
Newton and Wigner noted that the NW position operator is 
unique and the same as the mass-center position operator of Pryce (e) for a massive spin $1/2$ particle \cite{NW49}. 
The NW (Pryce (e)) position operator for a massive spin $1/2$ particle also satisfy all desirable commutation 
relations such as 
$[q^i_{\scriptsize{(e)}}, P^j]=i \delta_{ij}$ and $[q^i_{\scriptsize{(e)}},S^j_{\scriptsize{(e)}}]=0$.
This suggests that the spin operator ${\bf S}_{\scriptsize{(e)}} $ can be a unique relativistic spin 
operator for a massive spin $1/2$ particle.

%Foldy and Woutheysen have shown that the Pryce (e) position and spin operators have the usual form 
%${\bf x}$ and $\boldsymbol{\sigma}/2$ in the new equivalent representation, 
%in which the Hamiltonian is diagonalized. 

\section{Foldy-Woutheysen (FW) Spin Operator}

Foldy and Woutheysen \cite{FW50} have obtained the equivalent representation where the Dirac Hamiltonian 
changes into the diagonal form as 
\begin{eqnarray}
\mathcal{H}_{\scriptsize{FW}}=\beta E = U_{\scriptsize{FW}} \mathcal{H}_{\scriptsize{D}} U^{-1}_{\scriptsize{FW}},
\end{eqnarray}
by using a canonical transformation, 
\begin{eqnarray}
U_{\scriptsize{FW}} =\frac{E+m+\beta \boldsymbol{\alpha}\cdot {\bf P}}{\sqrt{2E(m+E)}}.
\end{eqnarray}
The representation is called the FW representation where the Hamiltonian is $\mathcal{H}_{\scriptsize{FW}}$. 
Foldy and Woutheysen considered the usual position operator ${\bf x}$ 
and the 4-dimensional Pauli spin operator $\boldsymbol{\Sigma}/2$ in the FW representation 
and have shown that the inverse canonical 
transformation gives the mean position operator ${\bf x}_{\scriptsize{FW}}$ and the mean 
spin operator ${\bf S}_{\scriptsize{FW}}$ in the standard Dirac representation as 
\begin{eqnarray}
{\bf x}_{\scriptsize{FW}}= U^{-1}_{\scriptsize{FW}} {\bf x} U_{\scriptsize{FW}}, ~~~
{\bf S}_{\scriptsize{FW}}= U^{-1}_{\scriptsize{FW}} \frac{\boldsymbol{\sigma}}{2} U_{\scriptsize{FW}}.
\end{eqnarray}
The position operator ${\bf x}_{\scriptsize{FW}}$ and the spin operator ${\bf S}_{\scriptsize{FW}}$ 
are guaranteed to satisfy the desirable commutation relations such as 
$[ { x}^i_{\scriptsize{FW}}, {x}^j_{\scriptsize{FW}}]=0$, $[ { \bf x}_{\scriptsize{FW}}, {\bf S}_{\scriptsize{FW}}]=0$, 
$[ { S}^i_{\scriptsize{FW}}, {S}^j_{\scriptsize{FW}}]=\ \epsilon_{ijk} S^k$, and 
$[{\bf S}_{\scriptsize{FW}}, \mathcal{H}_{\scriptsize{D}}]=0$  
because $[x^i, x^j]=0$, $[{\bf x}, \boldsymbol{\Sigma}]=0$, $[\Sigma^i, \Sigma^j]=2i \epsilon_{ijk}\Sigma^k$, and 
$[\mathcal{H}_{\scriptsize{FW}}, \boldsymbol{\sigma}]=0$ in the FW representation. 
They noted that the spin operator ${\bf S}_{\scriptsize{FW}}$ is the same 
as the spin operator of Pryce (e) \cite{FW50}.

In fact, Pryce also has defined the unitary operator $U_{\scriptsize{P}} $ transforming 
the usual position operator into ${\bf q}_{\scriptsize{(e)}}=  U^{-1}_{\scriptsize{P}} {\bf x} U_{\scriptsize{P}}$ 
in the standard Dirac representation, where 
\begin{eqnarray}
U_{\scriptsize{P}}= \frac{\boldsymbol{\alpha}\cdot {\bf p}+ \beta(E+m)}{\sqrt{2E(E+m)}}.
\end{eqnarray} 
The unitary operator is just $U_{\scriptsize{P}}=\beta U_{\scriptsize{FW}}$. 
It can be easily checked that the Hamiltonian, the position, and the spin operators in the standard Dirac representation 
obtained from the unitary similarity transformations $U_{\scriptsize{P}} $ are the same as the corresponding operators 
in the standard representation given by $U_{\scriptsize{FW}}$. 
The Pryce (e) position operator in the standard Dirac representation is the unique local position operator 
for massive spin $1/2$ particle as noted by Newton and Wigner \cite{NW49}. 
Hence the spin operator $ {\bf S}_{\scriptsize{FW}}$ in the standard Dirac representation 
is also the unique spin operator and becomes the usual 4-dimensional Pauli spin operator 
in the FW representation in an arbitrary frame. 

This fact, however, does not imply that the spin operators ${\bf S}_{\scriptsize{BG}}$ or ${\bf S}_{\scriptsize{NW}}$ 
can represent the spin operator 
in an arbitrary frame in FW representation. 
Note that the spin operator ${\bf S}_{\scriptsize{NW}}$ in Eq. (\ref{eq:NW}) has the same form as 
the spin operator ${\bf S}_{\scriptsize{BG}}$ in Eq. (\ref{eq:SPINBg}). 
The spin operators ${\bf S}_{\scriptsize{BG}}$ and 
${\bf S}_{\scriptsize{NW}}$ cannot be the usual Pauli spin operator in the FW representation. 
As was shown in Sec. \ref{sec:Pryce}, the spin operators ${\bf S}_{\scriptsize{BG}}$ and 
${\bf S}_{\scriptsize{NW}}$ become the usual 4-dimensional Pauli spin operator in the standard representation. 
Therefore, the transformed operator to the FW representation becomes 
$U_{\scriptsize{FW}} {\boldsymbol{\sigma}}/{2} U^{-1}_{\scriptsize{FW}}$ that is definitely 
not the Pauli spin operator. 

As was shown in Ref. \cite{Choi13}, the FW spin operator is equivalent to the covariant spin operator for 
a spin $1/2$ massive particle. The covariant spin operator is the same as the relativistic spin operator 
for a massive spin $1/2$ particle, 
which is recently derived from the first principle of space-time symmetry \cite{Choi14}. 
In Ref. \cite{Choi14}, a unique covariant spin operator has been derived from the general linear combination 
of PL vector. This unique covariant spin operator has been shown to be equivalent to the 
unique Pryce (e) (FW) spin operator. 
Therefore, the equivalence of the derived spin in Ref. \cite{Choi13} and the FW (Pryce (e)) spin implies that the physical properties 
related with the relativistic position and the spin operators for a massive relativistic particle with spin $1/2$ 
can be studied either covariantly or similar to non-relativistic quantum mechanics. 
In Ref. \cite{Choi11}, the entanglement change of the spin under the Lorentz transformation has been studied by using 
the FW spin operator and was shown to be equal to the results of Gingrich et. al. \cite{Gingrich02} obtained 
by using the Wigner little group method \cite{Wigner39}.  

\section{Summary}
\label{sec:SUM}
In summary, we have investigated three historically important position and spin operators proposed by Bogolubov et al., 
Pryce, and Foldy-Woutheysen to manifest a corresponding spin operator to the Newton-Wigner position operator 
for a unique localized state of a massive particle with spin. The spin operator of Bogolubov et al. 
is shown to be just the 4-dimensional Pauli spin operator that represents the spin operator in the particle rest frame. 
We have also shown that the Pryce (c) and (e) position and spin operators have the proper physical meaning when they are defined 
by using the non-covariant Hamiltonian for a relativistic massive particle with spin 1/2. 
However, the generalized definitions using covariant energy-momentum are shown to be different from the original definitions of Pryce 
because of the difference between the non-covariant Hamiltonian and the covariant 0th-component of momentum. 
We have explicitly reminded that the Pryce (e) position and spin operators are the same as the FW position and spin operators 
in the standard Dirac representation. The recently derived spin operator from the first principle of the Poincar´e group 
has been discussed to be equivalent as the spin the FW and Pryce (e) spin operator. 
These facts confirm that FW and Pryce (e) spin operator is the unique operator valid 
in relativistic quantum mechanics for a massive spin-1/2 particle.

\begin{center}
\section*{ACKNOWLEDGMENTS}
\end{center}
The authors are grateful for helpful discussions with Prof. Sam Young Cho at Chongqing University. 
This work was supported by a research grant from Seoul Women's University(2014).

\end{document}